\documentclass[12pt]{article}

\usepackage{amsmath}
\usepackage{graphicx}
\usepackage{mathrsfs}

\begin{document}
\def\scri{\unitlength=1.00mm
\thinlines
\begin{picture}(3.5,2.5)(3,3.8)
\put(4.9,5.12){\makebox(0,0)[cc]{$\cal J$}}
\bezier{20}(6.27,5.87)(3.93,4.60)(4.23,5.73)
\end{picture}}
\frenchspacing

\begin{center}

{\large BLACK HOLES: THEIR LARGE INTERIORS}

\vspace{10mm}

{\large Ingemar Bengtsson}

\

{\large Emma Jakobsson}

\vspace{7mm}

{\sl Stockholms Universitet, AlbaNova\\
Fysikum\\
S-106 91 Stockholm, Sweden}

\vspace{5mm}

{\bf Abstract:}

\end{center}

{\small 

\noindent Christodoulou and Rovelli have remarked on the large interiors 
possessed by static black holes. We amplify their remarks, and extend 
them to the spinning case.}

\vspace{8mm}

\begin{figure}[t]
	\begin{center}
	\leavevmode
	\includegraphics[width=0.9\textwidth]{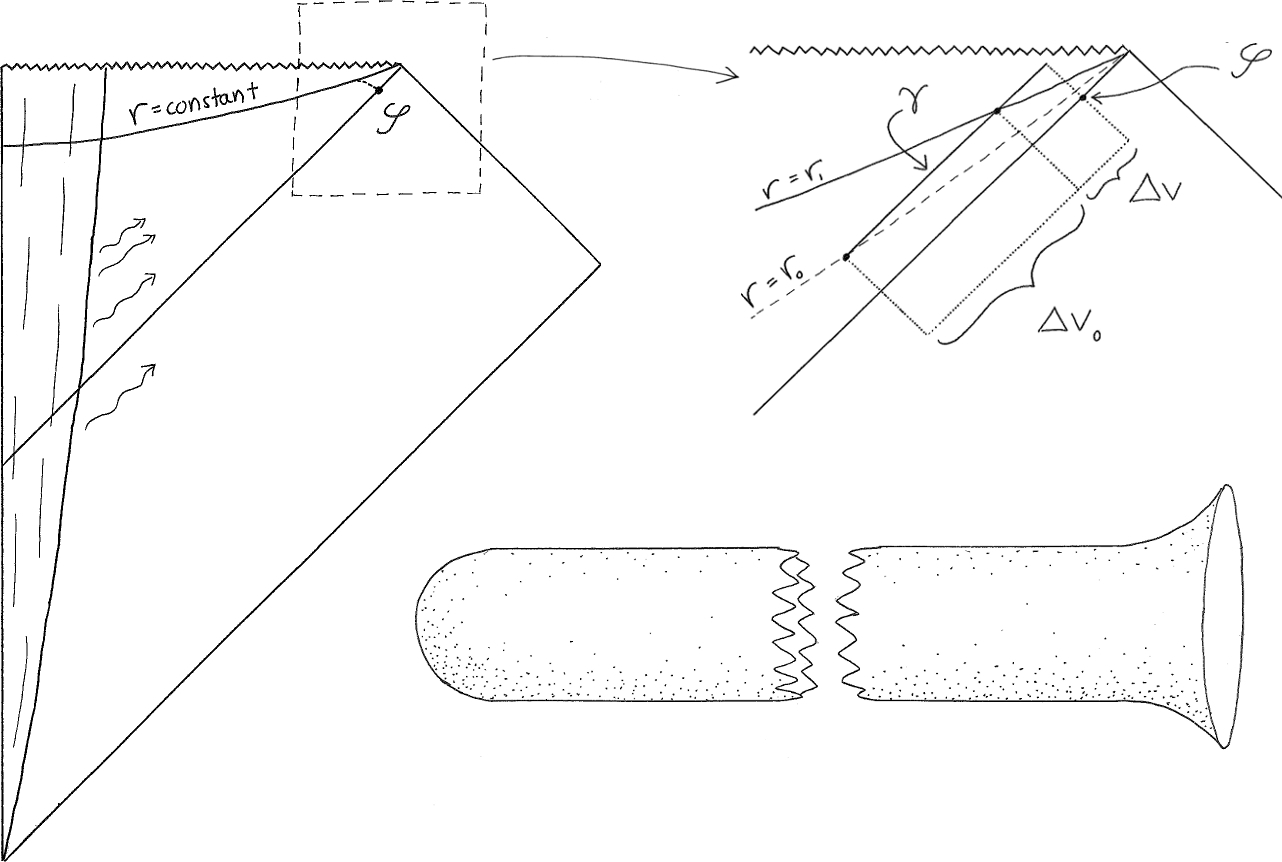}
	\end{center}
	\caption{\small A sphere $\mathscr{S}$ on the event horizon bounds a spacelike hypersurface, a large portion of which coincides with an $r =$ constant hypersurface. We show this hypersurface with one dimension suppressed, and cut in the middle, omitting the long cylindrical part which gives the main contribution to its volume. We also illustrate the argument showing that most of the volume is contained in a region out of causal contact with matter that has advanced far into the black hole.}
	\label{fig:2}
\end{figure}

\noindent The usual picture of a black hole is that of a compact object which 
eventually, after aeons of Hawking radiation, shrinks to a point and then 
disappears without a trace. But it is possible for a black hole to have a very 
large interior, in which case this picture is uncomfortably counterintuitive \cite{Ong}. 
Recently Christodoulou and Rovelli (CR from now on) \cite{CR} pointed out that 
black holes always have very large interiors. More precisely they pointed out 
that a cross section of the event horizon of a spherically symmetric black hole 
taken at late times (much later than the disappearance of the collapsing matter) 
bounds a spatial volume which grows with time as 

\begin{equation} \mbox{Vol} \sim 3\sqrt{3}\pi m^2 v \ , \label{1} \end{equation}

\noindent where $m$ is the mass of the black hole and $v$ is the advanced time. 
A look at the Penrose diagram (Fig. \ref{fig:2}) explains why. Consider a sphere 
$\mathscr{S}$ in the event horizon 
at large advanced time. (We assume that the metric is known to the reader. If not, 
see below.) A spacelike sphere bounds many 
spacelike hypersurfaces \cite{DiNunno}. Among them, choose one which is ``close to null'' just inside 
the sphere, then joins an $r =$ constant hypersurface all the way down to the 
matter filled region, and is closed off there. In the Schwarzschild region the $r=$ constant hypersurfaces 
are actually cylinders of constant radius. The Killing vector 
field $\partial_v$ acts along them. They contribute to the volume through the integral 

\begin{equation} \mbox{Vol} = \int^v dvd\theta d\phi \sqrt{2m/r-1}\ r^2
\sin{\theta} \ . \end{equation}

\noindent The lower integration limit is irrelevant since the integral 
will be dominated by its upper limit $v$. The smaller we choose $r$, the larger 
will be the first factor in the integrand---the cylinders are stretched. On the 
other hand their spherical cross sections will shrink. The coefficient in front 
of $v$ is maximized by choosing $r = 3m/2$, which yields eq. (\ref{1}). The contribution 
from the part of the hypersurface in the matter region is small---at late 
advanced time the leading term is always contributed 
by the Schwarzschild region. 

We must keep in mind that deviations from spherical 
symmetry are likely to have a large effect on the interior. For one thing, the 
singularity will no longer be Kasner-like (exhibiting stretching in one consistent 
direction, and contraction in the two others). Will this affect the result? 
We think not. If the mass of the black hole is large tidal forces will be 
small at the horizon, but they will also be rather small at $r = 3m/2$. In 
fact we are rather far from the singularity. Most of the volume is collected 
at late advanced time, from a region which is out of causal contact 
with that part of the matter region which has entered significantly into 
the black hole. 

Let us make the last part of the argument a bit more quantitative.
Recall that the advanced time is defined as $v = t + r_*$, and the retarded time as $u = t - r_*$, where
\begin{equation} r_*(r) = \int^r\frac{dr}{1-2m/r} = r + 2m\ln{(1-r/2m)} \ . 
\end{equation}
Consider a radial null geodesic $\gamma$ at constant $u$ extending from $r = r_0$ to $r = r_1$. It will cover an 
amount of advanced time equal to 

\begin{equation} \Delta v_0 = 2r_*(r_1) - 2r_*(r_0) \ . \label{eq3} \end{equation}

\noindent Let us assume that the event horizon has a radius of $10^6$ km. If 
we set $r_1 = 3m/2$ and $\Delta v_0 = 10^3$ years, we find that 

\begin{equation} \frac{c \Delta v_0}{\frac{2 G m}{c^2}} \approx 10^{10} \quad \Rightarrow \quad r_0 \approx 2\left(1-e^{-10^{10}}\right) m \ . \label{eq4} \end{equation}

\noindent From Fig. \ref{fig:2} it is clear that we have identified a region extending only very marginally indeed into the black hole, such that the main contribution to the volume (the term proportional to $\Delta v$) is contained in its Cauchy development.

If this argument is accepted it seems reasonable to repeat the calculation 
for a black hole that settles down to the Kerr black hole at late advanced time. 
If this were not the case the observation by CR would lose its bite, since 
real black holes do rotate, more or less.  
We expect that at large (but finite) advanced time the black hole will be 
close to Kerr not only outside the event horizon, but also in a part of the 
interior which is close to it. Thus we are interested in a region whose metric 
is closely approximated by \cite{Kerr}
%
\begin{eqnarray} ds^2 = - \frac{\Delta - a^2\sin^2{\theta}}{\rho^2} dv^2 + 2dvdr + 
\rho^2d\theta^2 + \frac{A\sin^2{\theta}}{\rho^2}d\phi^2 - \nonumber \\ 
\\
- 2a\sin^2{\theta}drd\phi - \frac{4amr}{\rho^2}\sin^2{\theta} dvd\phi \ , 
\hspace{22mm} \nonumber \end{eqnarray} 

\noindent where 
\begin{equation} \Delta \equiv r^2 - 2mr + a^2  \ , \end{equation}

\begin{equation} \rho^2 \equiv r^2 + a^2\cos^2{\theta} 
 \ , \hspace{5mm} 
A \equiv (r^2 + a^2)^2 - a^2\Delta\sin^2{\theta} 
\ . \end{equation}

\noindent The angular momentum is $J = am$. The event horizon is at $r = r_+$, which is the largest root of $\Delta = 0$. 
Radially ingoing null geodesics have constant advanced time $v$, and 
the coordinate $r$ has a physical interpretation as an affine parameter along 
these geodesics. 

We can now proceed just as in Schwarzschild: We start with a 
sphere on the event horizon at late advanced time, connect it to a spacelike 
hypersurface at constant $r$ in the interior, and close it once we reach values 
of $v$ where the geometry starts to deviate from the above. The main contribution 
to the volume of this hypersurface will be 

\begin{eqnarray} \mbox{Vol} = \int^vdvd\theta d\phi \sqrt{-\Delta}\rho \sin{\theta} 
= \hspace{33mm} \nonumber \\ \label{eq:vol} \\ = 2\pi v\sqrt{-\Delta}\left( 
\sqrt{r^2+a^2} + \frac{r^2}{2a}\ln{\frac{\sqrt{r^2+a^2}+a}{\sqrt{r^2+a^2}-a}}\right) 
\ . \nonumber  \end{eqnarray}
 
\noindent In the extreme limit $a/m = 1$, the region where $\Delta < 0$ disappears, and there is no 
such term. 

\begin{figure}
	\begin{center}
	\leavevmode
	\includegraphics[width=8cm]{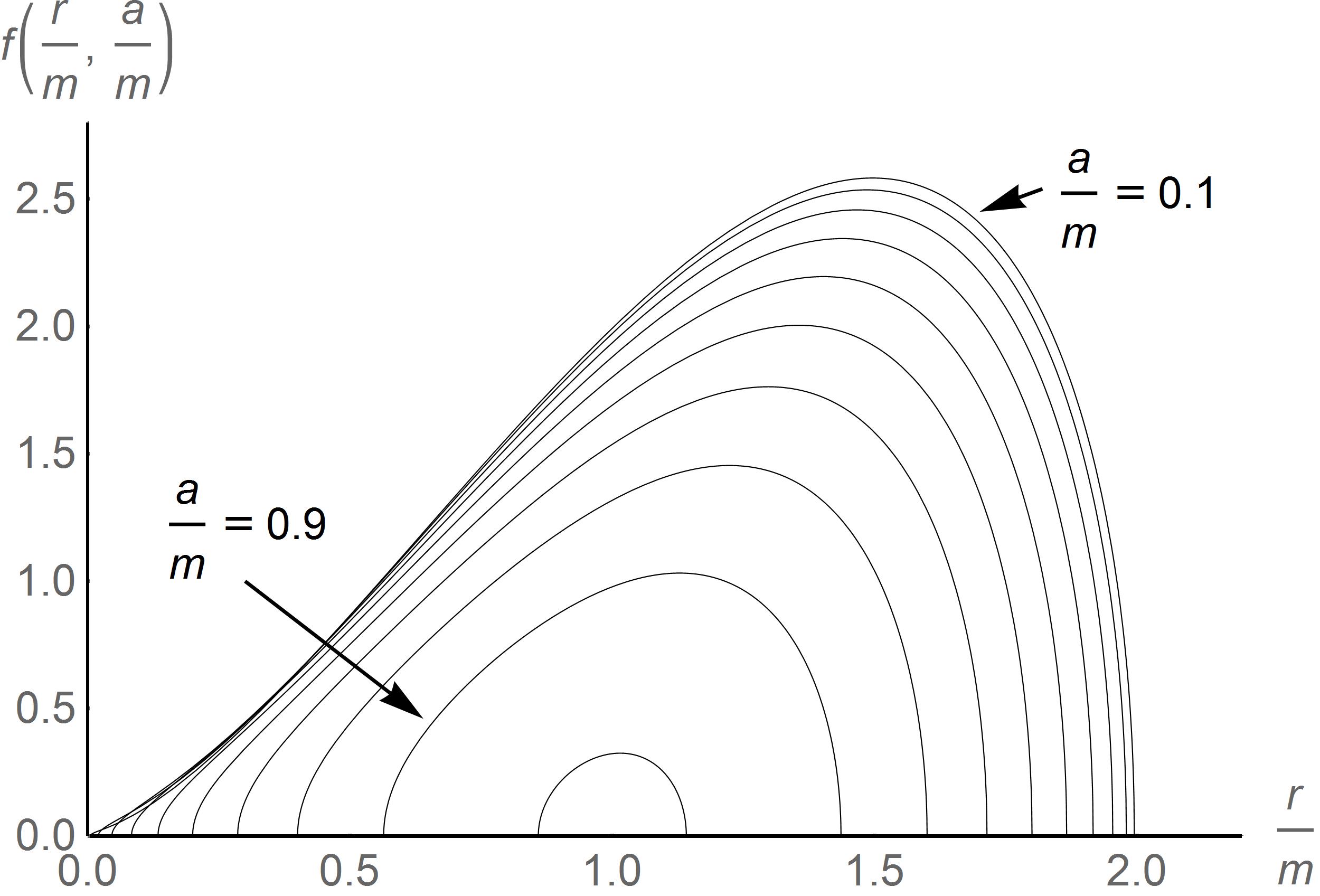}
	\end{center}
	\caption{\small The volume of a hypersurface of constant $r$ in the Kerr geometry is given by Vol $= 2\pi m^2 v f(\frac{r}{m},\frac{a}{m})$ (eq. \eqref{eq:vol}). The figure shows how the numerical factor $f$ depends on our choice of $r$, for values of $a/m$ ranging from 0.1 to 0.9 in even steps, and for $a/m = 0.99$ on the innermost curve. }
	\label{fig:1}
\end{figure}

It would make no sense to maximize this expression analytically, instead we give 
the results as Fig. \ref{fig:1}. What we see is that the value of the numerical 
coefficient in front of $v$ at maximal volume decreases with $a/m$, but the 
effect is not dramatic as long as $a/m < 0.99$. Thus we conclude that large 
volumes are present for realistic values of $a/m$. Of course the Kerr metric 
will fail to give a good approximation of the situation throughout much of 
the interior of the black hole. But if its exterior becomes indistinguishable 
from the Kerr black hole at $v = v_0$, then we only have to assume that this 
is so also for a very thin shell inside the event horizon. If we wait for 
another $10^3$ years the region where we perform the calculation will be 
out of causal contact with the interior of that shell. We convince ourselves 
of this by looking at radial null geodesics, just as we did for Schwarzschild. 
With ``radial'' we now mean that they belong to the outgoing Kerr congruence. 
Then eq. (\ref{eq3}) still holds, but with the modification that 

\begin{equation} r_*(r) = 
r + \frac{2m}{r_+-r_-}\left(r_+\ln{(1-r/r_+)} - 
r_-\ln{(r/r_- - 1)}\right) \ , \end{equation}

\noindent where $r_\pm$ are the roots of $\Delta = 0$ (radii of the outer and 
inner horizons). Then, if we start at a value of $r$ which maximizes the volume, and go back a thousand years in time, we find that eq. \eqref{eq4} generalizes to
\begin{equation}
	r_0 \approx \left( 1 - e^{-\frac{r_+ - r_-}{r_+} 10^{10}} \right) r_+.
\end{equation}
It remains true that 
the main contribution to the volume comes 
from the Cauchy development of a hypersurface that extends only marginally into 
the interior of the black hole. 
And we are all the time avoiding the neighbourhood of the inner horizon, where 
instabilities are likely to pile up. We conclude that the reasoning by 
CR survives the generalization to realistic spinning black holes. 

In their paper CR estimate that the black hole at the centre of the Milky Way---whose 
area radius they assume to be not much larger than the distance to the Moon---now 
contains enough space to fit a million solar systems. A decent estimate 
for its spin appears to be $a/m \approx 0.9$ \cite{Moscib}. It follows that 
CR overestimate the volume, but only by a factor of 10 or somewhat less. 

There have been a number of recent attempts to define the volume of a black hole 
\cite{Parikh, Gibbons, Lake, Dolan}. There is considerable freedom here, but we think 
that the observation made by CR is a striking one. In their paper they actually 
solve a kind of 
isoperimetric problem: Given a round sphere on the event horizon of a spherically 
symmetric black hole created by a collapsing null shell, what is the volume of 
the largest spherically symmetric 
hypersurface bounded by the sphere? The answer is as given above \cite{CR}.
In fact it has to be, because $r=3m/2$ defines a maximal hypersurface in Schwarzschild \cite{Reinhart,DeWitt}.
In the Kerr case hypersurfaces of constant $r$ are never maximal, so we do not expect our result to be optimal, but it may be nearly so.

\

\noindent \underline{Acknowledgement}: We thank Brandon DiNunno for a useful comment.

\end{document}